\documentclass[twocolumn,aps,floatfix]{revtex4}
\usepackage{graphicx}
\begin{document}
\title{Radiation back-reaction in relativistically strong
and QED-strong laser fields} 
\author{Igor V. Sokolov}
\email{igorsok@umich.edu}
\affiliation{Space Physics Research Laboratory, University of Michigan, Ann
Arbor, MI 48109 }
\author{John A. Nees}
\affiliation{Center for Ultrafast Optical Science and FOCUS Center, University
of Michigan, Ann Arbor, MI 48109}
\author{Victor P. Yanovsky}
\affiliation{Center for Ultrafast Optical Science and FOCUS Center, University
of Michigan, Ann Arbor, MI 48109}
\author{Natalia M. Naumova}
\affiliation{Laboratoire d'Optique Appliqu\'{e}e, 
ENSTA, Ecole Polytechnique,
CNRS, 91761 Palaiseau, France}
\author{G\'{e}rard A. Mourou}
\affiliation{
Institut de la Lumi\`{e}re Extr\^{e}me, 
ENSTA, Ecole Polytechnique,
CNRS, 91761 Palaiseau, France}
\date{\today}
\begin{abstract}
The emission from an electron in the field of a relativistically strong 
laser pulse is analyzed. At the pulse intensities of 
$\ge 10^{22}\ {\rm W/cm^2}$ the 
emission from counter-propagating electrons is modified by the 
effects of Quantum ElectroDynamics (QED), as long as the electron energy is 
sufficiently high: ${\cal E}\ge 1\ {\rm GeV}$. The radiation 
force experienced by an electron is for the first time derived from the QED principles 
and its applicability range is extended towards the QED-strong
fields.
\end{abstract}
\pacs{
52.38.-r Laser-plasma interactions,
41.60.-m Radiation by moving charges, 
52.38.Ph X-ray, gamma-ray, and particle generation }
\keywords{Thomson-Compton effect, Lorentz-Abraham-Dirac equation, 
radiation back-reaction}
\maketitle
In Quantum ElectroDynamics (QED) an electric field should be treated as  
strong if it exceeds the Schwinger limit: $E_S=m_ec^2/(|e|\lambdabar_C)$ \cite{schw}. 
Such
field is potentially capable of separating a virtual 
electron-positron pair providing an energy, which exceeds the electron rest 
mass energy, $m_ec^2$, to a charge, $e$, over an acceleration length as small 
as the Compton wavelength, $\lambdabar_C=\frac\hbar{m_ec}$.  
Particularly, a
QED-strong Lorentz-transformed electric 
field, $E_{0}=|{\bf p}\times{\bf B}|/(m_ec)$, may be exerted by  
a charged particle with the momentum ${\bf p}$, gyrating in the magnetic field, ${\bf B}$, 
if $|{\bf p}|\gg m_ec$ and$/$or the magnetic field is strong enough. 

Consider QED-effects in  
a strong pulsed laser field \cite{Mark}:  
\begin{equation}\label{eq:strong}
\sqrt{{\bf a}^2}\gg1,\qquad{\bf a}=\frac{e{\bf A}}{m_ec^2},
\end{equation} 
${\bf A}$ being the vector potential of the wave. 
In the laboratory frame of reference the electric field is not QED-strong for 
achieved laser intensities, $\sim10^{22}\ {\rm W/cm^2}$ \cite{1022}, and 
even for the $\sim10^{25}\ {\rm W/cm^2}$ intensity projected  \cite{ELI}. 
Nonetheless, a counter-propagating particle in a 1D  wave, ${\bf
a}(\xi),\,\xi=\omega t-({\bf k}\cdot{\bf x})$, may experience a QED-strong
field,  
$E_0=|d{\bf A}/d\xi|\omega({\cal E}-p_\|)/c$, because the laser frequency, $\omega$, is Doppler
upshifted in the frame of reference comoving with the electron. Herewith the electron dimensionless energy, 
${\cal E}$, and
its momentum are related to $m_ec^2$, and $m_ec$ correspondingly,
and subscript $\|$ herewith denotes the vector projection on the direction of the
wave propagation.  The Lorentz-transformed field exceeds the Schwinger limit, 
if 
$
\chi\sim E_0/E_S
=\frac{\lambdabar_C}{\lambdabar}({\cal
 E}-p_\|)\left|\frac{d{\bf a}}{d\xi}\right|\gg1$, 
where $\lambdabar=\frac c\omega$. 

Within classical theory a 
signature 
for $E_0$ is a radiation 
loss rate,  
$I_{\rm cl}(E_0)=2e^4E^2_0/(3m_e^2c^3)$. Therefore,  
the QED-strength of the 
electromagnetic
field may be determined in  
evaluating $I_{\rm cl}$ and  
its ratio to 
$I_C=I_{\rm cl}(2E_S/3)$: 
\begin{equation}\label{eq:chifirst}
\chi= \sqrt{\frac{I_{\rm cl}}{I_C}},\qquad
I_C=\frac{8e^2c}{27\lambdabar_C^2}.
\end{equation}
If $\chi\ge 1$ then  
the actual radiation loss rate differs 
from $I_{\rm cl}$.     
The 
condition of $\chi>1$ also separates 
the parameter range of the Compton effect from that of the Thomson effect, 
under the condition of Eq.(\ref{eq:strong}). 
The distinctive feature of the Compton effect is  
an electron  
recoil, which 
is significant, 
if a typical  
emitted photon energy, $\hbar\omega_c$, is comparable with the electron energy 
\cite{kogaetal}. Their ratio, $\chi=\lambdabar_C\omega_c/(c{\cal E})$, 
equals $\chi$ as defined in Eq.(\ref{eq:chifirst}) with the proper numerical factors 
(cf Eq.(\ref{eq:omegac})). 

Counter-propagating electrons can be generated in the course of laser pulse 
interaction with a solid target, that is why the 
radiation effects in the course of laser-plasma interaction are widely 
investigated (see \cite{kogaetal,lau03}). The principle matter in this paper is an account 
of the radiation back-reaction acting on a charged particle. This can be  
consistently done 
by solving the modified
Lorentz-Abraham-Dirac equation  
as derived in \cite{our},
in which the radiation back-reaction on the electron motion is expressed in 
terms of the emission probability. The calculation of this probability in
relativistically strong and QED-strong laser pulses is given in Section I. 
In Section II we discuss the radiation effect on the 
electron motion in strong fields.
\section{Electron in a 1D wave: the emission probability} 
The emission probability in the strong 1D wave field may be found in 
\S\S40,90,101 in \cite{lp}, as well as in \cite{nr},\cite{gs}. However, to 
simulate highly dynamical effects in pulsed fields, one needs 
a reformulated emission probability, related to short time intervals 
(not $(-\infty,+\infty)$).

{\bf Consider the classical motion of an electron}  in a 1D field, 
$a=a^\mu(\xi),\,\xi=(k\cdot x),\,(k\cdot a)=0$, $a$, $k$ and $x$ being the
4-vectors of the 
potential, the wave and the coordinates. Herewith
the 4-dot-product is introduced in a 
usual manner: $(k\cdot x)=\omega t-({\bf k}\cdot{\bf x})$ etc.,
3-vectors in contrast with 4-vectors being denoted in bold,
4-indices are omitted below.   
Introduce
a 
Transformed Space-Time (TST) :  
$x^{0,1}=(ct\mp x_\|)/\sqrt{2}$, $x^{2,3}={\bf x}_\perp$, subscript $\perp$ 
denoting the vector components 
orthogonal to ${\bf k}$. 
Note, that: (1)
$dx^0=\lambdabar d\xi/\sqrt{2}$, $p^0=\lambdabar(k\cdot p)/\sqrt{2}$, $(p\cdot k)= 
({\cal E}-p_\|)/\lambdabar$; 
(2) the 
momentum components,  
$p^0$ and ${\bf p}_\perp+{\bf a}$,  
are conserved; (3) the metric tensor in the TST is:
$
G^{01}=G^{10}=1,\,G^{22}=G^{33}=-1$; and (4)  
$p^\mu G_{\mu \nu}p^\nu=1$ gives: $p^1=(1+{\bf p}_\perp^2)/(2p^0)$.  
These properties allow us to find: 
\begin{equation}\label{eq:classicintense}
I_{\rm cl}=-\frac{2e^2c}{3}\frac{dp_j}{ds}G^{jk}\frac{dp_k}{ds}=\frac{2e^2c(k\cdot
 p)^2}{3}\left(\frac{d{\bf a}}{d\xi}\right)^2,
\end{equation}
and to relate  
4-momenta  
at different time instants:
\begin{equation}\label{eq:classic}
p(\xi)=p(\xi^\prime)-\delta a+
\frac{2(p(\xi^\prime)\cdot \delta a)-(\delta a)^2}{2(k\cdot p)}k,
\end{equation}
where $ds=\sqrt{c^2dt^2-({\bf p}\cdot d{\bf x}/{\cal E})^2}$ 
and $\delta a=a(\xi)-a(\xi^\prime)$. 
\begin{figure}
\includegraphics[scale=0.4, angle=90]{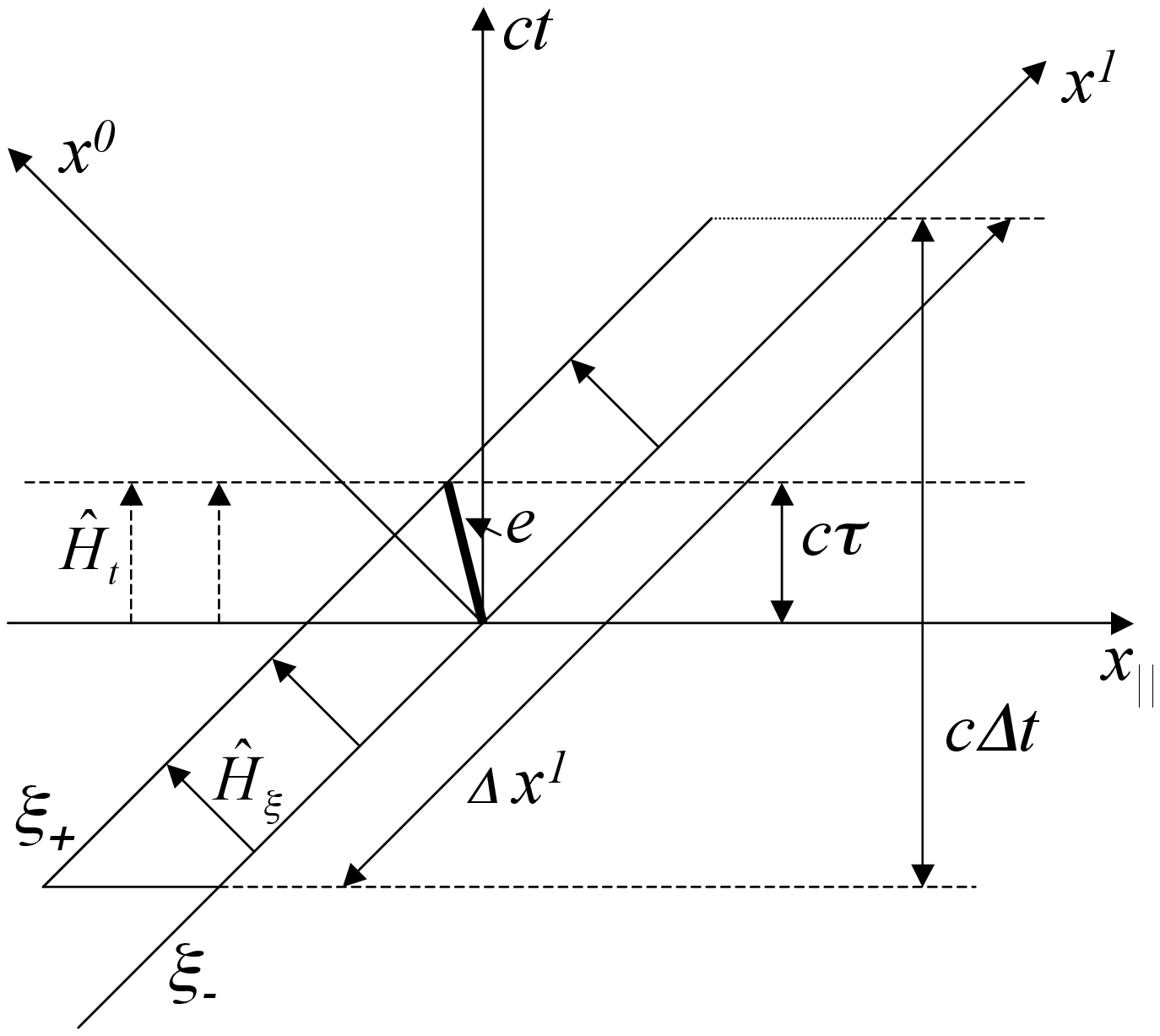} 
\caption{The volume over which to integrate the matrix element while finding 
the emission probability: in the standard scheme for the dipole emission 
(in dashed lines) and in the TST (in solid lines). Arrows show the 
direction, 
along which the Heisenberg operator advances the wave functions.
}
\label{fig_1}
\end{figure}  

{\bf A QED solution of the Dirac equation}  
in the TST is given by a plane electron wave ($N={\rm const}$),
\begin{equation}\label{eq:volkov}
\psi=\frac{ u(p(\xi))}{\sqrt{N}}C(\xi)
 \exp\left[\frac{i({\bf p}_{\perp0}\cdot {\bf x}_\perp)}{\lambdabar_C}
-\frac{i(k\cdot p)x^1}{(\lambdabar_C/\lambdabar)\sqrt{2}}\right].
\end{equation}
The Dirac equation,  
$
\left((\gamma\cdot(i \lambdabar_C\frac\partial{\partial x}-a))-1\right)\psi=0$, 
is satisfied 
under the following conditions: 
$u(p(\xi))$ is plane wave bi-spinor amplitude, 
$\left((\gamma\cdot p(\xi))-1\right)u(p(\xi))=0$, $\hat{u}\cdot u=2$, 
$\gamma^\mu$ are the Dirac matrices, and 
$C(\xi)=\exp\left(-\frac{i}{\lambdabar_C}
\int^{\xi}_{\xi^\prime}{\frac{1+{\bf p}_\perp^2(\xi_1)}{2(k\cdot p)}
d\xi_1}\right)
$. 
Using Eq.(\ref{eq:classic}), we find:
\begin{equation}\label{eq:QEDpropagator}
u(p(\xi))=\left[1+\frac{(\gamma\cdot
 k)\left(\gamma\cdot 
\delta a
\right)}{2(k\cdot p)}\right]u(p(\xi^\prime)).
\end{equation}  
 
{\bf The emission probability.} Introduce domain, $\Delta^4x=(\Delta x^1*S_\perp)*\Delta x^0$, 
bounded by two hypersurfaces, $\xi=\xi_{-}$ and $\xi=\xi_+$
(see Fig.1). The difference $\xi_+-\xi_-$ 
is bounded as decribed below,
so that $\Delta^4x$ covers only a minor part of the pulse.   
A volume $V=S_\perp\lambdabar(\xi_+-\xi_-)$, is a section
of $\Delta^4x$
subtended by a line $t={\rm const}$. With the 
choice of the coefficient in Eq.(\ref{eq:volkov}),
$
N=2S_\perp \lambdabar\int_{\xi_-}^{\xi_+}{{\cal E}
(\xi_2)d\xi_2}$,  
the integral 
$S_\perp \lambdabar\int_{\xi_-}^{\xi_+}{\hat{\psi}\gamma^0\psi d\xi_1}$ is 
set to unity, 
i.e. there is a single electron in the volume $V$. 
The emission probability,  $dW$, for a photon of 
wave vector, $k^\prime$, polarization vector, $l$, and wave function, $
A^\prime=
\frac{\exp[-i(k^\prime\cdot x)]}
{N_p^{1/2}}l$, $N_p=\frac{\omega^\prime V}{2\pi\hbar c^2}$, 
is given by an integral  
over $\Delta^4x$:
\begin{equation}\label{eq:probab}
dW=\frac{\alpha L_fL_p}{\hbar c}
\left|\int{\hat{\psi}_f(\gamma\cdot (A^\prime)^*)\psi_idx^0dx^1dx^2dx^3}\right|^2,
\end{equation} 
$L_p=
d^2{\bf k}^\prime_\perp d(k\cdot
k^\prime)\hbar c N_p/((2\pi)^2(k\cdot k^\prime))$ is the number of 
states for 
the emitted photon, a subscript $i,f$ denotes the 
electron 
in the intitial (i) or final (f) state, and $\alpha=\frac{e^2}{\hbar c}\approx\frac1{137}$. The 
number of electron states, $L_{i,f}$, per $d^3{\bf p}_{i,f}=d^2{\bf p}_{\perp
i,f}d(k\cdot p)_{i,f}{\cal E}_{i,f}(\xi)/(k\cdot p)_{i,f}$, in the wave field 
can be found by calculating the Hamiltonian 
invariant, 
$d^3{\bf p}dV\sim {\cal E}(\xi)d\xi$, 
with no field: 
$
L_{i,f}=
\frac{d(k\cdot p)_{i,f}d^2{\bf p}_{\perp i,f}N_{i,f}}
{2(2\pi)^3\lambdabar_C^3(k\cdot p)_{i,f}}.
$

{\bf Conservation laws.} The integration by $dx^1dx^2dx^3=c\sqrt{2} dtd^2{\bf x}_\perp$  
results in three $\delta-$ functions, expressing the conservation of 
totals of ${\bf p}_\perp$ and $(k\cdot
p)$, for particles in initial and 
final states. 
Twice integrated with respect to $dx^1$, the probability $dW$ is  
proportional to a 
long time interval, $\Delta t=\Delta x^1/(c\sqrt{2})$,  
if
the boundary condition for the electron wave 
at $\xi=\xi_-$
is maintained within that long time. For a single 
electron, which locates  between the wave fronts $\xi=\xi_-$ and $\xi=\xi_+$ during 
a shorter time, 
\begin{equation}\label{eq:time}
\delta t(\xi_-,\xi_+)=(1/c)\int_{\xi_-}^{\xi^+}{{\cal E}_i(\xi)d\xi_2}/(k\cdot p_i), 
\end{equation}
the emission probability is: $dW_{fi}(\xi_-,\xi_+)=\delta t dW/\Delta t$.  
Using $\delta-$ functions
we integrate Eq.(\ref{eq:probab}) over $d{\bf p}_{\perp f}d(k\cdot
p_f)$:
\begin{equation}\label{eq:abssq}
\frac{dW_{fi}(\xi_-,\xi_+)}{d(k\cdot k^\prime)d^2{\bf k}^\prime_\perp}=
\frac{\alpha \left|
   \int_{\xi_-}^{\xi_+}{F(\xi)\hat{u}(p_f)(\gamma\cdot l^*)u(p_i)d\xi
       }\right|^2}{(4\pi)^2(k\cdot k^\prime)(k\cdot p_i)(k\cdot p_f)},
\end{equation}
where 
$
F(\xi)=\exp[i
\int^\xi{(k^\prime\cdot p_{i}(\xi_2))d\xi_2}/
{(k\cdot p_{f})}]
$ and 
\begin{equation}\label{eq:cons}
p_f(\xi)=p_i(\xi)-\lambdabar_C k^\prime+\frac{(k^\prime\cdot p_i(\xi))\lambdabar_C}
{(k\cdot p_i)-(k\cdot k^\prime)\lambdabar_C}k . 
\end{equation}

{\bf To integrate Eq.(\ref{eq:abssq}),} we re-write it as the 
double integral over $d\xi d\xi_1$ and reduce the matrices 
$u(p_{i,f}(\xi))\otimes \hat{u}(p_{i,f}(\xi_1))$ in the integrand to the polarization 
matrices of the electron at $\xi$ or at $\xi_1$ using 
Eq.(\ref{eq:QEDpropagator}). Although in a strong wave 
electrons may be polarized (see \cite{Omori}),  
we then average over electron and photon polarizations and find: 
$$
\frac{dW_{fi}}{d(k\cdot k^\prime)d^2{\bf k}^\prime_\perp}=
\frac{\alpha \int_{\xi_-}^{\xi_+} {\int_{\xi_-}^{\xi_+}{F(\xi)F(-\xi_1)]D(\xi,\xi_1)
d\xi d\xi_1}}}{(2\pi)^2
(k\cdot k^\prime)(k\cdot p_i)(k\cdot p_f)},
$$ 
where $
D=-
\left( 
\frac{\left[{\bf a}(\xi)-{\bf a}(\xi_1)\right]^2\left((k\cdot p_i)^2+(k\cdot p_f)^2\right)
}{4 (k\cdot p_i)(k\cdot p_f)}+1\right).
$ 
Now we develop the dot-product, $(k^\prime\cdot p_i)$, in 
$F(\xi)$ in the TST  metric $G$ and find: 
$F(\xi)F(-\xi_1)=\exp\left[i(F_1+F_2)\right]$, where 
$$F_1=\frac{(k\cdot p_i)}{2(k\cdot k^\prime)(k\cdot p_f)}\left(\frac{(k\cdot k^\prime)}{(k\cdot p_i)}\left<{\bf p}_{\perp i}\right>-
{\bf k}^\prime_{\perp}\right)^2(\xi-\xi^*),
$$
$
F_2=\frac{ (k\cdot k^\prime) \left\{(\xi-\xi_1)+\int_{\xi_1}^\xi{
\left[{\bf a}(\xi_2)-\left<{\bf a}\right>\right]^2d\xi_2}\right\}}{2(k\cdot p_i)(k\cdot p_f)}
$, $\left<{\bf a}\right>=\frac{\int_{\xi_1}^\xi{{\bf a}d\xi_2}}{\xi-\xi_1}$  and 
\begin{equation}
\frac{dW_{fi}(\xi_-,\xi_+)}{d(k\cdot k^\prime)}=
\frac{\alpha \int_{\xi_-}^{\xi_+} {\int_{\xi_-}^{\xi_+}{\frac{i\exp(iF_2)}{\xi-\xi_1}D(\xi,\xi_1)
d\xi d\xi_1}}}{2\pi(k\cdot p_i)^2}.
\end{equation} 

{\bf In the strong field as in Eq.(\ref{eq:strong})} the formulae simplify.
In $F_2$ we estimate:   
$\xi-\xi_1\sim
\left|d{\bf a}/d\xi\right|^{-1}$,  and
$(k\cdot k^\prime)\sim (k\cdot p_i)(k\cdot p_i)\left|d{\bf a}/d\xi\right|$. 
Now we can {\it conistently} introduce the bounds for $\xi_+-\xi_-$: 
$$\left|d{\bf a}/d\xi\right|^{-1}\ll\xi_+-\xi_-\ll
\min\left(\alpha^{-1}\left|d{\bf a}/d\xi\right|^{-1},1\right).
$$
Under these bounds, 
the emission probability: (1) is linear in $\xi_+-\xi_-$: $dW_{fi}(\xi_-,\xi_+)=(dW_{fi}/d\xi)(\xi_+-\xi_-)$; 
(2) is less than unity: $\int{dW_{fi}}<1$; and (3) can be expressed in terms of the local 
electric field. By introducing
$\left<\xi\right>=(\xi+\xi_1)/2$,
$\theta=(\xi-\xi_1)|d{\bf a}/d\xi|/2$, 
$({\bf a}(\xi)-{\bf a}(\xi_1))^2\approx 4\theta^2$ 
and expressing the integral over $\theta$  
in terms of the MacDonald functions we find:
\begin{equation}\label{eq:probabf}
\frac{dW_{fi}}{dr_0d\xi}=
\frac{\alpha \chi\left(\int_{r_\chi}^\infty{K_{\frac53}(y)dy}+r_0r_\chi\chi^2
K_{\frac23}(r_\chi)\right)}{\sqrt{3}\pi(k\cdot p_i)\lambdabar_C},
\end{equation}
$$
r_\chi=\frac{r_0}{1-\chi r_0},\,\,\,\,
\chi=\frac32(k\cdot p_i)\left|\frac{d{\bf a}}{d\xi}\right|\lambdabar_C=\sqrt{\frac{I_{\rm cl}}{I_C}},
$$
Probability (similar to that found in \cite{nr}) is expressed in terms of functions of $r_0=\frac23\frac{(k\cdot k^\prime)}{(k\cdot p_i)^2|d{\bf a}/d\xi|}$, and 
related to interval of
$dr_0$. Below we demonstrate the way to use this probability to describe the electron 
motion and emission.
\section{Radiation and its back-reaction}
\begin{figure}
\includegraphics[scale=0.38]{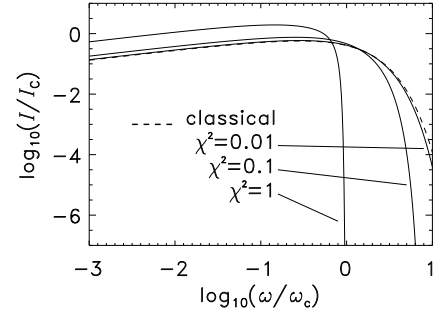}
\caption{Emission spectra for various values of $\chi$. 
}
\label{fig_2}
\end{figure}
\begin{figure}
\includegraphics[scale=0.38]{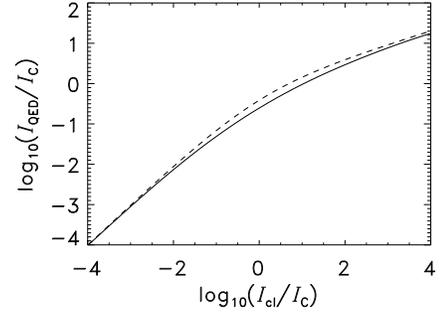}
\caption{Emitted radiation power in the QED approach {\it vs} classical 
(solid); an interpolation formula $I_{\rm QED}=I_{\rm cl}/(1+1.04\sqrt{I_{\rm cl}/I_C})^{4/3}$ (dashed).}
\label{fig_3}
\end{figure}
At ${\bf p}_\perp^2\gg1$  
the wave vector of the emitted photon 
is almost parallel to the electron momentum.   
For 
colinear $k^\prime$ and $p_i$, one has
${\omega^\prime}/{{\cal E}_i}\approx c{(k\cdot k^\prime)}/{(k\cdot p_i)}$,
therefore,
\begin{equation}\label{eq:omegac}
r_0=\frac{\omega^\prime}
{\omega_c},\,\,\,\,\omega_c=\frac{3}{2}{\cal E}c(k\cdot p_i)\left|\frac{d{\bf a}}{d\xi}\right|={\cal E}c\chi/\lambdabar_C.
\end{equation}
The assumption, ${\bf p}_\perp^2\gg1$ also allows us to find 
the momentum of the emitted radiation, which we relate to the interval of the electron 
{\it proper} time, using Eqs.(\ref{eq:time},\ref{eq:probabf}):
\begin{eqnarray}
\frac{dp_{\rm rad}}{d\tau}=
\int{\lambdabar_C k^\prime \frac{c(k\cdot p_i)dW}{ d(k\cdot k^\prime)d^2{\bf k}_\perp d\xi}d(k\cdot k^\prime)d^2{\bf k}_\perp}=\nonumber\\
=[p+k\ O((k\cdot p_i)^{-1})]\int{c\lambdabar_C (k^\prime\cdot k) \frac{dW}{dr_0 d\xi}dr_0}.\label{eq:prad}
\end{eqnarray}
As with other 4-momenta, $p_{\rm rad}$ is related to $m_ec$. To prove the 4-vector relationship (\ref{eq:prad}), we expand 
its components in the TST metric and integrate them over ${\bf k}_\perp$ using the symmetry of $F_1$. The small term,
$O(1/(kp_i))$, arises from the electron rest mass energy  
and from the small 
($\sim 1/|p_\perp|$) but finite width of the photon angular distribution. 
Below we neglect
this term 
and find: $dp_{\rm rad}/d\tau=p_i\frac1{m_ec^2} (dI_{\rm QED}/dr_0)dr_0$
where $I_{\rm QED}=m_ec^2\int{c\lambdabar_C (k\dot k^\prime)\frac{dW}{d\xi dr_0}dr_0}$ is the radiation loss rate.

Thus, the angular distribution can be represented as $\delta({\bf \Omega}-{\bf p}/\sqrt{{\bf
p}^2})d{\bf \Omega}$,  with ${\bf
\Omega}$ being the solid angle of the photon direction. 
The photon energy spectrum, $dI_{\rm QED}/dr_0$,  is described as function only of
the random {\it scalar}, $r_0=\omega^\prime/\omega_c$, using only the parameter, $\chi$ (see Fig.\ref{fig_2}). 
The latter may be parameterized in terms of the radiation loss rate, evaluated within 
the framework of classical theory (see Eq.(\ref{eq:chifirst}) and Fig.\ref{fig_3}). The 
expressions for $q(I_{\rm cl})=I_{\rm QED}/I_{\rm cl}$ and for the 
normalized spectrum function, $Q(r_0,\chi)$,  coincide with formulae known from the 
gyrosynchrotron emission theory (see \S90 in \cite{lp}): 
$$
q=\frac{9
\sqrt{3}}{8\pi}\int_0^\infty{dr_0r_0\left(\int_{r_\chi}^\infty{K_{\frac53}(y)dy}+r_0r_\chi\chi^2
K_{\frac23}(r_\chi)\right)},
$$
$$
Q(r_0,\chi)=\frac{9\sqrt{3} r_0}{8\pi q}\left(\int_{r_\chi}^\infty{K_{\frac53}(y)dy}+
r_0r_\chi\chi^2
K_{\frac23}(r_\chi)\right)
$$
{\bf Radiation back-reaction.} While emitting a photon, an electron also acquires 
4-momentum from the external field, equal to $dp_F=\frac{(k^\prime\cdot p_i(\xi))\lambdabar_C}
{(k\cdot p_i)-(k\cdot k^\prime)\lambdabar_C}k\approx k \lambdabar_C (k^\prime\cdot p_i)/(k\cdot p_i)$ (see Eq.(\ref{eq:cons})). 
Usually this is small compared to $dp_{\rm rad}$. However, the account for the interaction with the field ensures that 
the {\it total} effect of emission on the electron not to break the entity $(p_f\cdot p_f)=1$. The choices of near-unity correction 
coefficients in $dp_F$ are somewhat different in the cases $\chi\le 1$ and $\chi\gg 1$. For moderate values of $\chi$ the 
{\it radiation force}, $(dp_F-dp_{\rm rad})/dt$, 
may be found by integrating both $dp_F$ and $dp_{\rm rad}$ 
over $dk^\prime$:
\begin{equation}\label{eq:radf}
\frac{d(p_f-p_i)}{d\tau}=\left(k\frac{(p_i\cdot p_i)}{(k\cdot p_i)}-p_i\right)\frac{I_{\rm QED}}{m_ec^2},
\end{equation} 
where the choice of the coefficient in $dp_F$, first, ensures that the radiation force maintains the abovementioned entity 
(since $(p_i\cdot d(p_f-p_i)/d\tau)=0$), and, second, makes Eq.(\ref{eq:radf}) applicable with dimensional momenta as well. 

We already mentioned in \cite{our}, that QED is not compatible 
with the traditional approach to the radiation force in classical electrodynamics and 
suggested an alternative equation of motion for a radiating 
electron:
\begin{equation}\label{eq:our}
\frac{dp^i}{d\tau}=
\Omega^{ik}p_k-
\frac{I_{\rm QED}}{m_ec^2}p^i+\tau_0\frac{I_{\rm QED}}{I_{\rm cl}}
\Omega^{ik}\Omega_{kl}p^l,
\end{equation} 
where $\Omega^{ij}=eF^{ij}/(m_ec)$, $F^{ij}$ is the field tensor and $\tau_0=2e^2/(3m_ec^3)$. In the 1D plane wave
$\tau_0\Omega^{ik}\Omega_{kl}p^l=k^i(p\cdot p)I_{\rm cl}/(m_ec^2(k\cdot p))$, so that the radiation force in Eq.(\ref{eq:our})
is the same as its QED formulation in Eq.(\ref{eq:radf}). This proves that the 
earlier derived Eq.(
\ref{eq:our}) has a wide range of applicability including an electron quasi-classical 
motion in QED strong fields. The way to solve Eq.(\ref{eq:our}) and integrate the 
emission is described in 
\cite{our}.

\begin{figure}
\includegraphics[scale=0.4]{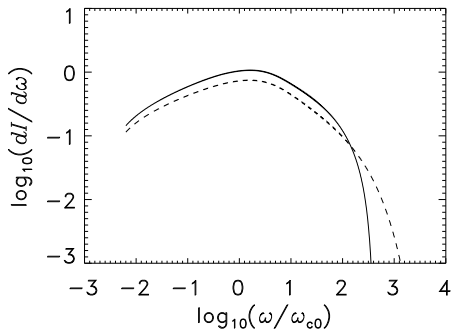}
\caption{The emission spectrum for 600 MeV electrons interacting with 
30-fs laser pulses of intensity $2\cdot 10^{22} W/cm^2 $: with (solid) or 
without (dashed) accounting for the QED effects. Here
$\hbar\omega_{c0}\approx 1.1$ MeV for $\lambda=0.8\mu$m.
}
\label{fig_5}
\end{figure}

In Fig.\ref{fig_5} we show the numerical result for an electron interacting with a
laser pulse. We see that the QED effects essentially modify the radiation spectrum even 
with laser intensities which are already achieved.  

{\bf We conclude} that in a wide range of applications, including the case of very 
strong laser fields with essential QED effects, the electron motion may be 
successfully described within the radiation force approximation. The necessary 
corrections in the radiation force and the emission spectra to account for the QED 
effects are parameterized by the {\it sole} parameter, $I_{\rm cl}$.

{\bf The future application to QED Monte-Carlo simulations} may be based on
the total probability of emission per interval of proper time:
$
W=\Delta\tau(I_{\rm QED}/(m_ec^2))\tilde{\omega}^{-1}$, where $\tilde{\omega}^{-1}=\int{Q(r_0,\chi)/(\chi r_0)dr_0}$. The expression of the only scalar to gamble,
$\lambdabar_C\omega^\prime/{\cal E} c$, in terms of a random number,
$0\le R<1$, is  
given 
by an integral equation as follows: 
$$
\int^{\lambdabar_C\omega^\prime/({\cal E} c\chi)}_0{Q/(r_0\chi)dr_0}=R<\lambdabar_C\omega^\prime/{\cal E}c>^{-1}.
$$
This method will be described in a forthcoming publication in detail, including the 
pair production (see \cite{kb} regarding the latter effect).

\end{document}